# Advancing Standards-Free Methods for the Identification of Small Molecules in Complex Samples

Jamie R. Nuñez[1], Sean M. Colby[1], Dennis G. Thomas[1], Malak M. Tfaily[1], Nikola Tolic[1], Elin M. Ulrich[2], Jon R. Sobus[2], Thomas O. Metz[1,*], Justin G. Teeguarden[1,3,*], Ryan S. Renslow[1,*]

[1]Earth and Biological Sciences Directorate, Pacific Northwest National Laboratory, Richland, WA, USA.

[2]U.S. Environmental Protection Agency, Office of Research and Development, National Exposure Research Laboratory, Research Triangle Park, NC, USA.

[3]Department of Environmental and Molecular Toxicology, Oregon State University, Corvallis, OR, US

\* ryan.renslow@pnnl.gov

**ABSTRACT:** The current gold standard for unambiguous identification in metabolomics analysis is based on comparing two or more orthogonal properties from the analysis of authentic, pure reference materials (standards) to experimental data acquired in the same laboratory with the same analytical methods. This represents a significant limitation for comprehensive chemical identification of small molecules in complex samples since this process is time-consuming and costly, and the majority of molecules are not yet represented by standards, leading to a need for standards-free identification. To address this need, we are advancing chemical property calculations and developing multi-attribute scoring and matching algorithms to utilize data from multiple analytical platforms through the utilization and creation of the *in silico* Chemical Library Engine (ISiCLE) and the Multi-Attribute Matching Engine (MAME). Here, we describe our results in a blinded analysis of synthetic chemical mixtures as part of the U.S. Environmental Protection Agency's (EPA) Non-Targeted Analysis Collaborative Trial (ENTACT). The blinded false negative rate (FNR), false discovery rate (FDR), and accuracy were 57%, 77%, and 91%, respectively. For high confidence identifications, the FDR was 35%. After unblinding of the sample compositions, we improved our approach by optimizing the scoring parameters used to increase confidence. The final FNR, FDR, and accuracy were 67%, 53%, and 96%, respectively. For high confidence identifications, the FDR was 10%. This study demonstrates that standards-free small molecule identification and multi-attribute matching methods can significantly reduce reliance on standards.

## INTRODUCTION

Conventional metabolomics and small molecule identification approaches have demonstrated immense value for disease diagnosis, evaluation of environmental exposures, and discovery of novel molecules. This success is reflected in the large number of recent biomedical research, environmental exposure studies, and soil and ecology publications employing metabolomics approaches.[2-10] In contrast to genetic and proteomic information available from rapid genome sequencing and proteome characterization, far less is understood about the totality of human exposure and small molecules found in the environment.[11-13] Furthermore, driven by a broader interest in understanding biological impacts of chemical exposures, biomonitoring is undergoing a significant evolution.[15,16] Traditional biomonitoring approaches, either targeted (seeking to identify specific compounds) or non-targeted (seeking to identify as many compounds as possible),[17] and using either low or high resolution mass spectrometry,[18,19] rely on authentic, pure reference materials (standards) for unambiguous chemical identification, and are therefore limited to the subset of molecules for which these standards exist.[20] A wealth of information about human exposure continues to emerge from these methods for a subset of chemical space confined to a priority list of molecules represented by standards. The Centers for Disease Control and Prevention (CDC) National Health and Nutrition Examination Survey (NHANES) program and the National Institutes of Health (NIH) Children's Health Exposure Analysis Resource (CHEAR) centers have provided such data, leading to examples of successful applications of these methods.[21-23]

Recent proposals to characterize the whole metabolome and exposome—the aggregate of all exposures—are driving a shift from traditional quantitative analytical chemistry and the typical strictures for chemical identification to new methods applicable to the discovery of molecules for which there are standards.[24] The vast chemical space of the metabolome and exposome together includes endogenous (e.g., molecular transducers and microbiomes) and exogenous (e.g., xenobiotics, industrial chemicals, consumer products, and transformation products of these) chemicals.[16] There are not enough authentic reference materials for the preponderance of these molecules. For example, using an automated script, we found only 17% of compounds found in the Human Metabolome Database, HMDB,[25] and less than 2% of compounds found in exposure chemical databases like the EPA DSSTox (comptox.epa.gov) can be purchased in pure form. Without chemical standards, unambiguous chemical identification is limited to the small number of molecules amenable to nuclear magnetic resonance spectroscopy- or crystallography-based structural elucidation while the vast majority is left as chemical "dark matter".[26] The need for more comprehensive and unambiguous chemical identification in these studies is driving innovations in analytical chemistry, computational chemistry and cheminformatics.[27,28] For example, new targeted and non-targeted methods have emerged as adaptions to traditional analytical chemistry.[29]

We are advancing standards-free metabolomics, the identification of small molecules without reliance upon standards, through the use of calculated chemical properties and associated matching using multiple experimental attributes (i.e., multi-attribute matching). Our approach currently relies on multiple experimental data types, including accurate mass, isotopic distribution, and collisional cross section (CCS), and comparison of these values to entries in *in silico* libraries, leveraging instrumental and computational innovations developed at Pacific Northwest National Laboratory (PNNL).[20,30-39]

To evaluate our methodology, we participated in the U.S. Environmental Protection Agency's (EPA) Non-Targeted Analysis Col-

laborative Trial (ENTACT), an inter-laboratory challenge established to provide a consistent set of verified, blinded synthetic mixtures for the objective testing of non-targeted analytical chemistry methods.[40,41] We performed blinded analysis of 10 synthetic mixtures each containing an, at the time, unknown number of substances as part of the multi-laboratory challenge. Accurate mass, isotopic signature, and CCS measurements were collected using ion-mobility spectrometry-mass spectrometry (IMS-MS) and ultra-high resolution 21-Tesla Fourier transform ion cyclotron resonance–mass spectrometry (FTICR-MS) (Figure 1). These properties were also calculated for each molecule in our processed form of the EPA Toxicity Forecaster (ToxCast)[42] library, allowing us to match observed features (e.g., peaks characterized by a measured mass and intensity, or a measured mass, CCS and intensity) to library entries. After unblinding, we performed statistical analysis on the results in order to find how well our method performed compared to others. Scoring parameters were then optimized to improve our method and to better understand the importance of each parameter in our scoring algorithm. Our findings demonstrate the potential of standards-free small molecule identification methods, particularly the value of using calculated, orthogonal properties, such as CCS and accurate mass, and multi-attribute matching to increase confidence in compound identification and significantly reduce reliance on standards.

## MATERIALS AND METHODS

**Blinded Small-Molecule-Spiked Samples.** Ten synthetic mixtures were provided by the EPA. Each mixture contained an unknown number of substances (later revealed as 95-365 substances) in dimethylsulfoxide (DMSO), with all substances selected from EPA's ToxCast chemical library[42]. Further details on ENTACT are outlined in Sobus et al. (2017)[41] and Ulrich et al. (2018).[40]

**Mass Spectrometry.** Briefly, samples were analyzed using both an Agilent 6560 drift tube ion mobility spectrometry-quadrupole time-of-flight mass spectrometer (IMS-MS)[30,43] and a 21-Tesla Fourier transform-ion cyclotron resonance spectrometer coupled to a Velos Pro dual linear quadrupole mass spectrometer (FTICR-MS)[31,32,44] in both positive (+) and negative (-) ionization modes (Figure 1). IMS-MS samples were analyzed using electrospray ionization (ESI) and atmospheric pressure photoionization (APPI). FTICR-MS samples were analyzed using ESI only. Samples were analyzed in triplicate using IMS-MS and in singlet using FTICR-MS. This resulted in 14 disparate experimental data sets per sample. Appropriate sample blanks provided by the EPA were also analyzed using each instrument method. Extensive details regarding the experimental protocol for sample preparation and both mass spectrometry methods are provided in Supporting Information (SI) 1.0-3.0.

**Chemical Property Calculations.** Mass, CCS, and isotopic signature were calculated for the $[M+H]^+$, $[M+Na]^+$, and $[M-H]^-$ adducts of each entry in the suspect library. We recently developed an automated high-accuracy method for calculating CCS and other chemical properties[20,33,39] called the *in silico* Chemical Library Engine (ISiCLE), only requiring chemical structure information (e.g., as provided by the InChI[45]).

The ISiCLE module for calculating IMS CCS for molecules has three methods for calculating CCS — *Standard*, *Lite*, and *AIMD-based* — of which the *Standard* and *Lite* methods were used for this study. Complete details regarding CCS calculation methods are provided in SI 4.0. At its current stage of development, the *Standard* method has an average error of 3.2% and the *Lite* method has an average error of 6.7% (SI Table 1);[46] however, the *Lite* method is much less computationally intensive, making it more than 200 times faster. The *Standard* method was used for calculating the CCS of all three adducts from a selected subset of 1,000 molecules that showed significant evidence (early in our analysis) of being present in the mixtures. The *Lite* method was then used for the remaining molecules, as an appropriate tradeoff between accuracy and computational cost based on the scope of the project. The CCS calculation method and results for each entry in the suspect library are provided in the Supplemental Data. Details on how we processed the ToxCast library to generate our suspect library are provided in SI 5.1.

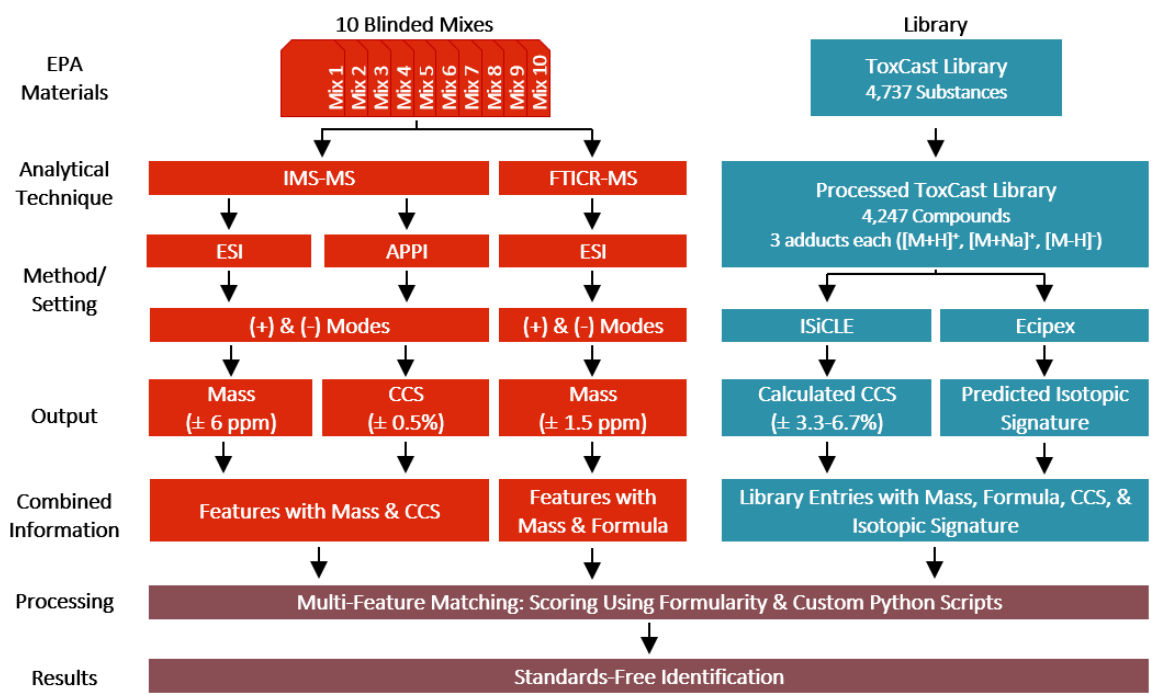

*Figure 1. Project overview. Detailed project flow, starting from the blinded mixtures and ToxCast Library (the given suspect screening library). After instrumental analysis of the mixtures and computational property calculations for library entries, our multi-attribute scoring algorithm was used for assigning confidence and identifying compounds likely to be present in each mixture. Note substances can be composed of one or more molecules that separate upon solvation in liquid. Molecules are single molecular structures.*

Ecipex[47] was used to calculate the isotopic signature of each adduct of each molecule in the suspect library. Once high-mass resolution data was collected using FTICR-MS, there was evidence for a significant presence of chlorinated compounds, leading to the additional calculation of chlorinated library entries (giving a total of four adducts per molecule with calculated isotopic signatures). Formularity[34] was then used to match calculated and observed isotopic signatures. Note, Formularity was not used on the IMS-MS data sets due to instrumental error being too high to reliably assign formulae to potential isotopic signatures. More details on how Formularity was used are provided in SI 3.2.

**Multi-attribute Downselection, Matching, and Scoring.** We developed a comprehensive identification package, the Multi-Attribute Matching Engine (MAME), which includes feature downselection and a scoring system to provide confidence scores (broken into low, medium, and high confidence). The confidence scores are increased by the number and quality of experimental features that match to those in the *in silico* library for a given entry and provide increasing evidence for the presence of the molecule in the sample. We scored the confidence of suspect library entry being in each mixture using our weighting method.

Note that the method described here does not label specific features as belonging to a specific compound (i.e., directly linking a feature arising due to instrument response to a specific compound), which is common in the literature. Instead, our scoring system considers all evidence indicating the presence of a compound, where multiple features consistent with possible instrument responses of a compound increase the probability of that compound's presence. The focus is to connect the experimental evidence to the presence of specific compounds, rather than attempt to prove that specific features resulted from specific compounds. This is an important distinction as it is not always possible to label a feature as belonging to a particular compound, especially in the case of complex samples. Instead, we use multiple experimental features to lend confidence to the presence of a compound within a sample, without attempting to label individual features.

Downselection of candidate features and molecular library entries and confidence scoring were performed using MAME, which processed all 14 disparate raw data sets per sample to achieve standards-free, multi-attribute, aggregate evidence-based molecular identification. A set of parameter cutoffs were used for data preprocessing (Table 1). For example, for an IMS-MS feature to be counted toward the confidence score of a molecule, it needed to (i) be observed in all three technical replicates, (ii) have a signal intensity $\geq 1000$ (arbitrary units), (iii) have a mass measurement error $\leq \pm 6$ ppm, and (iv) not have been observed in more than one blank (which also had three technical replicates). For an FTICR-MS feature to be counted toward the confidence score of a molecule, it needed to (i) have a mass measurement error $\leq \pm 1.5$ ppm, and (ii) not have been observed in the blank run. Initially, these cutoff criteria were chosen based on expert domain knowledge.

Once all analytical features were processed and matched to corresponding entries in the suspect library, we scored the confidence of each library entry being in each mixture using MAME, which uses a total of 11 independent scoring parameters (Table 2). These parameters were initially selected based on expert domain knowledge in our group, since this type of study had not been pursued previously. A library entry was labeled as "present" in the mixture if its confidence score was 6.0 or more. We decided evidence amounting to this score (e.g., observing a high intensity FTICR-MS feature (4 points) for a library entry with a unique mass (2 points)) was enough to earn this label. Confidence scores of 6.0-11.0, 11.0-19.0, and 19.0+ were labelled as low, medium, and high confidence, respectively. In addition, we apply the level system developed by Schymanski et al.,[48] which can be used to evaluate the level of confidence based on evidence provided by orthogonal features. Based on the given definitions, we use Level 2a to indicate identification based on mass and CCS, Level 4 for mass and isotopic signature, and Level 5 for identification based on mass alone. A more detailed description of MAME is included in SI 5.2-5.3 and the full software package is available upon request. As an example, Figure 2 shows how pioglitazone was scored and correctly labeled as present in one of the mixtures.

**Analysis and Optimization of our Scoring Algorithm.** As metrics to quantify success, we used false discovery rate (FDR, the percentage of false positives out of the total number of compounds labeled as present), false negative rate (FNR, also known as the miss rate, the percentage of false negatives out of how many compounds were spiked in by the EPA), and accuracy (the percentage of correct labels). Equations for each of these metrics are provided in SI 5.4. The overall goal of our method is to minimize FDR and FNR, while maximizing overall accuracy. When reporting these values here, we use the average across the ten mixtures. These metrics (and more) are broken down for each mixture in the Supplemental Data.

*Table 1. Parameter cutoffs used for scoring and downselection.*
[a] *Arbitrary units.* [b] *A library entry's mass is considered unique if its nearest neighbor library entry is more than 6 ppm away.*

| Category | Parameter | Cutoff |
|---|---|---|
| IMS-MS | Intensity | $\geq 1000$ a.u. [a] |
| | Mass Error (Magnitude) | $\leq \pm 6$ ppm |
| | # Seen in Samples | 3 |
| | # Seen in Blanks | $\leq 1$ |
| | CCS Error | $\leq 5\%$ |
| FTICR-MS | Intensity | $\geq 1$ a.u. |
| | Mass Error (Magnitude) | $\leq \pm 1.5$ ppm |
| | # Seen in Samples | 1 |
| | # Seen in Blanks | 0 |
| IMS-MS & FTICR-MS | High Intensity | $\geq$ 30th %ile |
| | Low Intensity | $<$ 30th %ile |
| Library | Unique Mass [b] | $> \pm 6$ ppm |
| | Large Mass | $\geq 200$ Da |

*Table 2. Initial scoring criteria and their associated weights.*
[a] *Earned a maximum of one time per library entry*

| Category | Index | Criteria | Weight |
|---|---|---|---|
| IMS-MS | 1 | High Intensity | 2.0 |
| | 2 | Low Intensity | 1.0 |
| | 3 | Low CCS Error | 3.0 |
| FTICR-MS | 4 | High Intensity | 4.0 |
| | 5 | Low Intensity | 2.0 |
| | 6 | Isotopic Signature | 3.0 [a] |
| IMS-MS & FTICR-MS | 7 | Additional Adducts | 1.0 |
| | 8 | Additional Features | 0.5 |
| | 9 | Detected by Both MS | 2.0 |
| Library | 10 | Unique Mass | 4.0 |
| | 11 | Large Mass | 1.0 |

It is important to note our analysis and statistics were based on identifying which structures were observed, not on identifying the correct parent compound. For example, cyclohexylamine and cyclohexylamine hydrochloride were both suspects provided in the ToxCast library. Since these are indistinguishable in solution, they were grouped into a single entry within our suspect library. We define successful identification of these compounds to mean we report both parent compounds as potential candidates when one or both was spiked into a mixture.

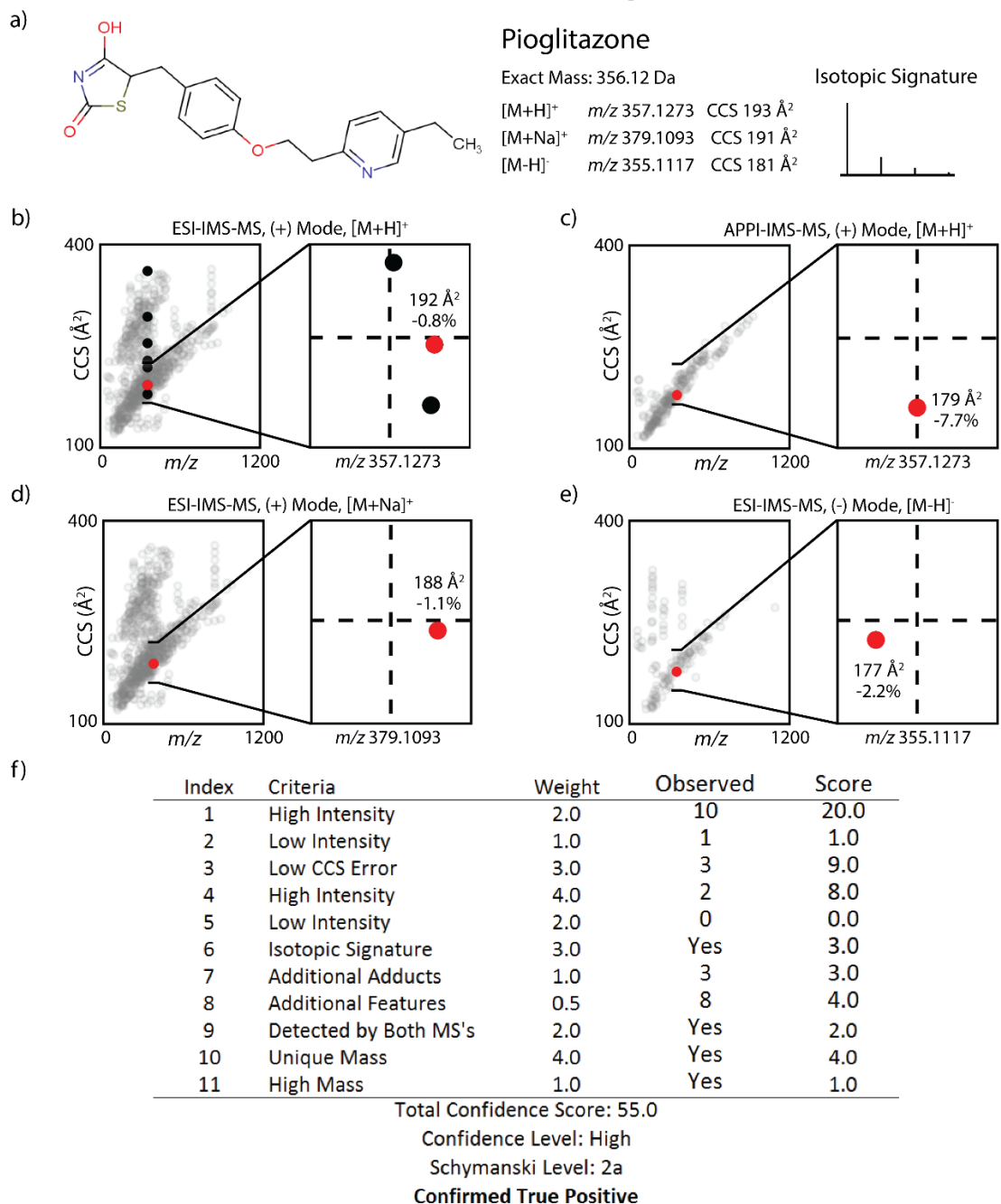

*Figure 2. Example scoring of pioglitazone, a true positive evaluated using our multi-attribute scoring system. Note, pioglitazone hydrochloride was in the ToxCast library, then changed to pioglitazone (the structure present in solution) in our processed library. a) Library entry for pioglitazone, a phenol ether[1] drug (sold as Actos) used to control high blood sugar in patients with type 2 diabetes,[14] with calculated CCS (using standard ISiCLE) for the three adduct types and its calculated isotopic signature. b-e) IMS-MS features observed within a ±6 ppm mass error window of a given adduct mass. A magnified view is provided, centered around the calculated mass and CCS, with the mass and CCS ranges extending 6 ppm and 20 Å, respectively, on either side of this average. Percentages are in respect to the calculated CCS. Red points indicate the experimental feature closest to our prediction. f) Combined scoring of all features. The number of features matching a specified criterion, or whether the criterion was met, is provided in the "Observed" column.*

After unblinding, we used Monte Carlo[49] and particle swarm optimization (PSO, via PySwarm[50]) methods, implemented in Python scripts, to select new weights for each scoring criteria using an objective function to maximize the area under the precision-recall curve (AUPR). AUPR is generated by determining precision and recall, which can be derived directly from FDR and FNR, respectively, parameterized by a minimum confidence score cutoff. This enables performance of the scoring weights to be assessed without an explicit cutoff selection for the score, which is a nontrivial decision with implications beyond the scope of this work.[51] AUPR was also selected as the objective function due to its relatively good performance compared to other classifiers when dealing with imbalanced datasets (i.e., significantly more true negatives compared to true positives).[52] Further details are provided in SI 5.5.

## RESULTS AND DISCUSSION

The foundation of our standards-free approach is the *in silico* construction of a library of chemical properties used to characterize experimental data collected for each sample. Our method operates by considering the consistency between the library of predicted properties and the observed analytical features, and subsequently quantifying and weighting their similarity. Calculated scores based on the evaluation of experimental features matched to library entries allow us to determine a single confidence score for each library entry and, ultimately, whether there is enough evidence to indicate a given compound is present in a sample.

**Construction of the Standards-Free *in silico* Library**. As part of this challenge, the EPA provided the ToxCast library as the suspect library (mixtures were only spiked with ToxCast substances). We processed all substances within this library as described in SI 5.1, which lead to a suspect library of 4,348 total compounds that are theoretically observable by mass spectrometry.

Approximately 50% of this library was not identifiable based on mass alone with an experimental mass error of ±6 ppm (Figure 3a). Further, 47% of library entries have at least one other formula conflict within the ToxCast and over 13% had five or more conflicts. Even perfect mass accuracy has high collision rates in nearly all chemical libraries, and thus high-resolution mass instruments alone are inadequate for high-accuracy identification without complementary, orthogonal data.[53]

CCS is a chemical property that provides additional information on which to increase the uniqueness of each library entry (Figure 3b). This is especially powerful when considering the CCS of each adduct as independent information, effectively adding corroborating dimensions of data for each adduct with a known CCS. Beyond the need to add additional dimensions beyond mass, CCS can also increase confidence of a molecule being present in the mixture by providing additional evidence and Schymanski Level 2a confidence rather than Level 4 or 5. Ultimately, the addition of CCS increased the confidence score of 90% of compounds that were correctly determined to be present in the samples.

**Experimental Analysis of Samples and Feature Extraction**. A total of 14 data sets were successfully generated per sample (plus additional data for blanks). The raw IMS-MS data included ~200,000 total *m/z*-CCS features observed across triplicate analyses and ~475,000 total *m/z* features observed across FTICR-MS analyses (Figures S1-2). This data showed evidence of many different adducts, as indicated by the high number of features and a significant presence of multimers, as indicated by frequent observations of features with extremely high CCS: *m/z* ratios (Figures S3-6 and see Figure 3b for the much tighter *m/z*-CCS distribution of the library when considering [M+H]+ features).[37] To help reduce noise and low-level contamination, we downselected to a subset of features using constraints based on feature intensity and presence across technical replicates (and absence in blanks), applying the cutoffs described in Table 1.

For IMS-MS, an intensity cutoff of 1,000 was set for all features in addition to requiring that each feature must have been observed across all three technical replicates and no more than once across the corresponding blank replicates. This removed 94% of features and improved confidence that those remaining were from molecules present in the sample. There was still significant evidence of multimer formation in our positive mode ESI-IMS-MS analyses (Figure S3b) but, with no way to ensure these were removed without losing possible overlapping monomer features, we decided to move forward, understanding that most of the suspected multimer features would not match the CCS values of library entries.

For FTICR-MS, we did not find a reliable method to apply an intensity cutoff, so the intensity cutoff was trivially set to 1. Since there was only a single replicate for each condition (due to limited sample), any feature seen at any intensity in the blank was removed, leading to 9% of features being removed.

**Blinded, Small Molecule Identification using Standards-Free Multi-Attribute Matching**. Before unblinding the true compositions of the mixtures, we performed multi-attribute matching by comparing the measured properties of downselected experimental features to our *in silico* library of calculated properties, and scoring each putative match using values given in Table 2. Note, for IMS-MS, the high intensity cutoff (i.e., the 30th percentile value of downselected features) was 2,123 and 2,174 for positive and negative mode, respectively. For FTICR-MS, the high intensity cutoff was 3,358 and 110 for positive and negative mode, respectively. An example of our multi-attribute scoring method is demonstrated in Figure 2.

This same analysis was performed for all library entries, taking into consideration all 14 disparate data sets (and blanks), using our Python module, MAME, resulting in a confidence score for each library entry for each mix. We submitted the list of compounds labeled as present in each mixture (and their associated confidence scores and confidence levels) to the EPA, who then unblinded the samples by returning the sample key to enable the assessment of our approach. An overview of the results is shown in Figure 4a-d.

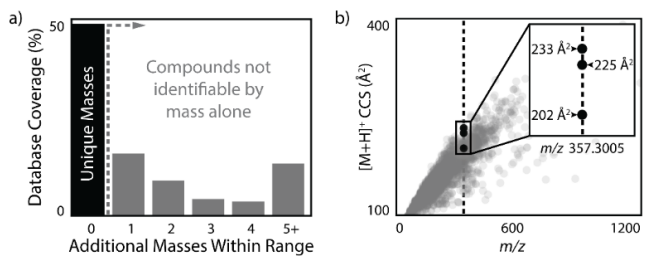

*Figure 3. CCS is a chemical property that increases each library entry's uniqueness. (a) Number of molecule entries in the ToxCast library (shown as a percentage of total library size) whose protonated masses fall within ±6 ppm of another entry. Zero (black bar) indicates no neighbors within this mass range (a molecule that can be resolved with mass alone, 2,216 total). The grey bars represent molecules (2,130 total) that cannot be distinguished based on mass alone, within an instrumental error of ±6 ppm. (b) Calculated CCS vs. m/z for the protonated forms of each molecule in the ToxCast library. The inset shows the example of m/z 357.3005, where 3 molecules lie within ±6 ppm of one another. When adding the property of CCS, all 3 molecules are predicted to become analytically unique within our specified parameter thresholds.*

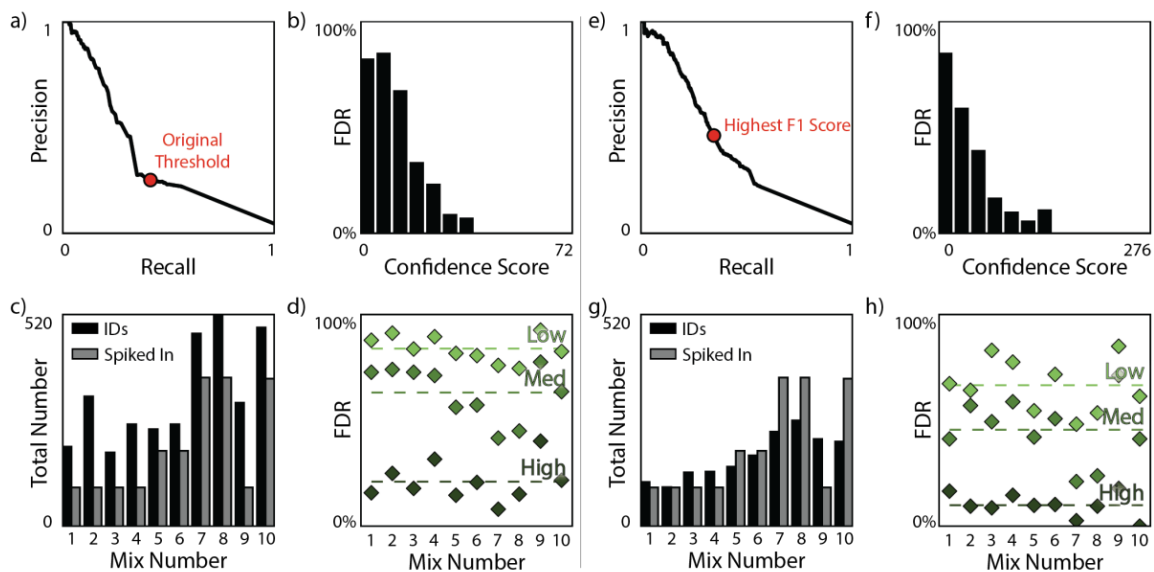

*Figure 4. Results using our standards-free multi-attribute matching methods. (a-d) Blinded method results. (e-h) Optimized (weights chosen using Monte Carlo) results. (a, e) AUPR curve, with red dot showing our cutoff (a total confidence score of 6.0 and 11.2 for blinded and optimized approach, respectively). Please refer to the SI for details on the highest F1 score. (b, f) FDR as a function of confidence score. (c, g) Comparison between the number of molecules identified compared to the number of molecules spiked into each mixture. (d, h) FDR for each of the mixtures individually, split by confidence levels.*

Our overall FDR, FNR, and accuracy was 77%, 57%, and 91%, respectively. For high confidence (confidence score of 19.0 or more) Schymanski Level 2a (probable structure provided by mass and CCS) identifications, FDR was 35%. Additionally, FDR had a smooth inverse trend with the magnitude of confidence score assigned by our algorithm (Figure 4b). We also showed the capability to distinguish between compounds with the same mass (including isomers) (Figure S7).

One major issue, which caused a high FDR, was that we routinely determined 300-500 molecules to be present in sample mixtures designed to contain 95-365 substances (Figure 4c). We hypothesized the high occurrence of false positives was attributable to one or more of the following factors: (i) noise present in raw data; (ii) low confidence score cutoff; (iii) detection of molecules that were in the suspect library, but unintentionally present in the samples due to reactions occurring in the highly concentrated mixtures; and/or (iv) multimer formation during the ionization process due to high sample concentrations. In the case of multimers, we hypothesized these formed during ESI, remained as multimers upon entry and flight through the IMS drift tube, and then dissociated to the constituent monomer prior to arriving at the MS detector.[37] Support for this hypothesis was provided by much higher observed CCS values than expected with corresponding *m/z* values that were consistent with monomers. Because our criteria for labeling a compound as present required associated experimental features to be observed across all three technical replicates (in the case of IMS-MS features), and minimal presence (observed once at the most) in blanks, it seems unlikely that low levels of contamination were the cause of the high FDR.

Chemical reactions that produce molecules found in the library, such as hydroxylation, are possible at the high concentrations found in the mixture. The EPA confirmed that each molecule was spiked in at approximately 0.05 mM. As a clear example, Figure 5 shows tamoxifen and 4-hydroxytamoxifen (a hydroxylated form of tamoxifen), molecules both found in the suspect library and both receiving high confidence scores (45.5 and 25, respectively) in the same sample. However, only tamoxifen was classified as a true positive since it was spiked into the mixture, whereas 4-hydroxtamoxifen was not. It is possible 4-hydroxytamoxifen may not be a genuine false positive and instead could have been formed in situ, due to reactions within the mixture.

Additionally, it is important to note the importance of choosing a cutoff (i.e., minimum score to be labeled as present) that best reflects the desired balance of true positives to true negatives. For example, during a forensics study, it may be desirable to decrease the number of false positives and therefore a higher cutoff would be needed. This would decrease FDR but also increase FNR. For example, in our case, increasing the cutoff from 6 to 19 leads to an FDR of 35%, FNR of 81%, and accuracy of 96%.

We then optimized the cutoff by finding the one that yielded the highest F1 score (a function of FNR and FDR, equation provided in SI 5.5). We found a cutoff of 9.5 (and using the same set of weights as our blinded approach) decreased FDR by 14% (to 63%), increased FNR by 9% (to 66%), and increased accuracy by 4% (to 95%) (Figure S8).

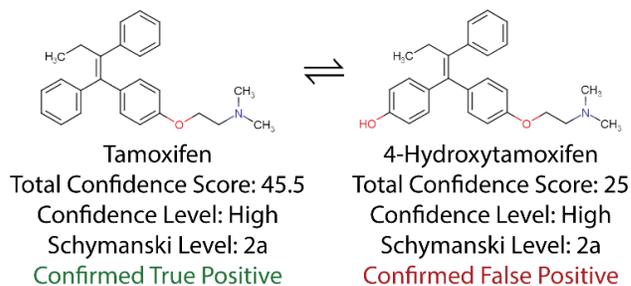

*Figure 5. Tamoxifen and 4-hydroxytamoxifen. Both were identified with high confidence in the same mixture but only tamoxifen was actually present.*

Based on these initial results, we concluded our approach worked well, but would likely be improved by optimizing our scoring parameters and cutoff ranges for each confidence category. Beyond finding which false positives were present due to the reasons stated earlier, this was the most powerful way to improve our overall results and learn more about our algorithm before broader application.

**Optimization of Standards-Free Multi-Attribute Matching Approach.** To determine the importance of each scoring parameter and to increase the accuracy of our approach, we set out to optimize our scoring method and subsequent confidence level cutoffs. The results of the Monte Carlo and particle swarm optimization methods are provided in the SI (SI 5.7, Figures S9-10). Optimization results were used to better understand the effect of each parameter and to update weights (Table S2), ultimately decreasing our combined FNR and FDR (Figure 4e-h).

## CONCLUSIONS

The capability to routinely measure and identify even a modest fraction of biologically, environmentally, or medically important chemicals within all of chemical space remains one of the grand challenges in science. The vast majority of molecules are not represented by standards. Furthermore, data for even fewer molecules have been added to reference libraries for use in identification (libraries currently cover much less than 1% of chemical space). This limit has remained a major constraint for decades in the global search for chemical biomarkers of disease, toxin exposure, and affiliated efforts in the search for new drug candidates and attempts to sequence the complete metabolome. It is clear that relying on a single instrument and slow, costly establishment of reference libraries in the laboratory, restricted to standards available for purchase, is not a viable approach for identifying the tens-to-hundreds of thousands of small molecules in complex biological or environmental samples. Through advances in instrumentation, computation, and data integration, there has been a push for a shift in metabolomics and exposomics toward standards-free, multi-attribute identification, in which the use of multiple molecular properties, accurately predicted computationally and consistently measured experimentally, are used for comprehensive identification of small molecules without the need for standards.

Our findings, both pre- and post-optimization, show great value in using standards-free, multi-attribute based identification methods. Furthermore, the addition of CCS increased confidence for true positives and was able to distinguish between isomers, even with our team's most rapid and least accurate CCS calculation method used for most molecules. To improve our results in the future, we will need to add additional capabilities that can be predicted or calculated.

This indicates the value for future use of additional identification "dimensions", such as MS/MS fragmentation patterns, chromatographic retention time, more accurate prediction of adduct formation (e.g., additional metal ion adducts not considered here), and infrared or Raman spectra. Complete standards-free identification, for even large library sizes, and potentially the complete molecular universe, may become possible through use of multiple accurately measured and calculated chemical properties. The value in increasing accuracy of analytical and computational methods is important; however, adding orthogonal chemical properties for all researchers in the field to use will aid in the identification of small molecules and will be essential for addressing major challenges within metabolomics. As additional chemical properties are added to this pipeline, the "distance" between the features of each library entry will become dramatically larger, thereby requiring a lower resolution for each property. The so-called "curse of dimensionality"[54,55] can be used for our benefit to turn each library entry into a unique or nearly-unique set of chemical properties with no overlapping neighbors. As metabolomics evolves and computational libraries are used more frequently, associated methods could eventually challenge the field's current definition of, and requirements for, identification.

While it is not possible to measure values such as accuracy in real (i.e., non-synthetic) complex mixtures, the approach described here was developed using blinded results. In future studies, we plan to again validate this approach using the optimized scoring parameters on other synthetic mixtures and real samples where molecules have already been identified with standards. Consistent low FNR, FDR, and accuracy with the same scoring system will show the use and reliability of our method in complex sample identification.

## ASSOCIATED CONTENT

**Supporting Information.** The Supporting Information is available free of charge on the ACS Publications website.

SupportingInformation.pdf: Includes further detailed methods and additional figures.

SupportingData.xlsx: Includes our suspect library, property predictions, and results broken down for each mixture in this challenge. Table captions are provided in SupportingInformation.pdf

## AUTHOR INFORMATION

**Corresponding Authors.**

Ryan S. Renslow - ryan.renslow@pnnl.gov
Justin G. Teeguarden - jt@pnnl.gov
Thomas O. Metz – thomas.metz@pnnl.gov

**Author Contributions.** The manuscript was written through contributions of all authors. All authors have given approval to the final version of the manuscript.

## ACKNOWLEDGMENTS

This research was partially supported by the Genomic Science Program (GSP), Office of Biological and Environmental Research (OBER), the U.S. Department of Energy (DOE), and is a contribution of the Pacific Northwest National Laboratory (PNNL) Metabolic and Spatial Interactions in Communities (MOSAIC) Scientific Focus Area (SFA). The Multi-Attribute Matching Engine (MAME) was fully developed under MOSAIC funding. Portions of this research were also supported by the United States Environmental Protection Agency (Interagency Agreement DW-089-92452001-0 in support of DOE Project No. 68955A), the National Cancer Institute (grant R03CA222443), and a PNNL Laboratory Directed Research and Development program, the Microbiomes in Transition (MinT) Initiative. This work was performed in the W. R. Wiley Environmental Molecular Sciences Laboratory (EMSL), a DOE national scientific user facility at the PNNL. The NWChem calculations were performed using the Cascade supercomputer at the EMSL. PNNL is operated by Battelle for the DOE under contract DE-AC05-76RL0 1830.

**For Table of Contents Only**

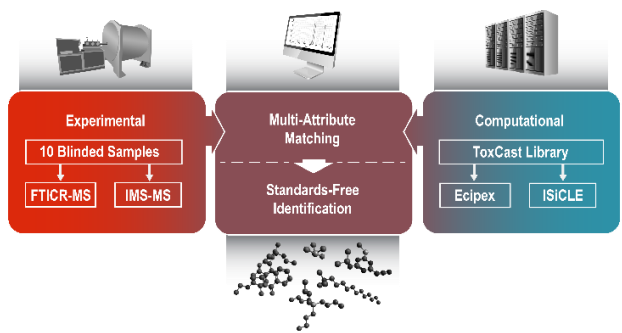